\documentclass[12pt]{article}
\usepackage {amsmath}
\usepackage[dvips]{graphicx}
\usepackage{feynmp}
\usepackage{amssymb}
\usepackage{cite}
\newcommand{\be}{\begin{equation}}
\newcommand{\ee}{\end{equation}}
\newcommand{\bear}{\begin{eqnarray}}
\newcommand{\eear}{\end{eqnarray}}

\newcommand{\lapproxeq}{\lower .7ex\hbox{$\;\stackrel{\textstyle  
<}{\sim}\;$}} 
\newcommand{\gapproxeq}{\lower .7ex\hbox{$\;\stackrel{\textstyle  
>}{\sim}\;$}} 
\newcommand{\stackdown}[2]{\lower 1.4ex\hbox{$\;\stackrel{\textstyle{#1}}  
{\scriptstyle{#2}}\;$}}
\newcommand{\beq}{\begin{equation}} 
\newcommand{\eeq}{\end{equation}} 

\newcommand{\ba}{\begin{eqnarray}}
\newcommand{\ea}{\end{eqnarray}}

\newcommand{\bea}{\begin{eqnarray}}
\newcommand{\eea}{\end{eqnarray}}

\makeatletter 
\def\slash{\@ifnextchar[{\fmsl@sh}{\fmsl@sh[0mu]}} 
\def\fmsl@sh[#1]#2{%
  \mathchoice 
    {\@fmsl@sh\displaystyle{#1}{#2}}%
    {\@fmsl@sh\textstyle{#1}{#2}}%
    {\@fmsl@sh\scriptstyle{#1}{#2}}%
    {\@fmsl@sh\scriptscriptstyle{#1}{#2}}} 
\def\@fmsl@sh#1#2#3{\m@th\ooalign{$\hfil#1\mkern#2/\hfil$\crcr$#1#3$}} 
\makeatother 
\begin{document}
\begin{titlepage}  
\begin{flushright} 
\parbox{4.6cm}{UA-NPPS/BSM-07/02 }
\end{flushright} 
\vspace*{5mm} 
\begin{center} 
{\large{\textbf {On the radion mediation of the Supersymmetry 
breaking in N=2, D=5 Supergravity Orbifolds.
}}}\\
\vspace{14mm} 
{\bf G. A.~\ Diamandis}, \, {\bf B. C.~\ Georgalas}, \, 
{\bf P.~\ Kouroumalou } and \\ {\bf A. B.~\ Lahanas}
{\footnote{email alahanas@phys.uoa.gr}}
\vspace*{6mm} \\
  {\it University of Athens, Physics Department,  
Nuclear and Particle Physics Section,\\  
GR--15771  Athens, Greece}

\end{center} 
\vspace*{25mm} 
\begin{abstract}
 We discuss the on-shell $N=1$ Supersymmetric coupling of brane chiral multiplets 
 in the context of $N=2$, $D=5$ Supergravity compactified on $S_1/Z_2$ orbifolds. 
 \newline
 Assuming a constant superpotential on the hidden brane we study the transmission 
 of the supersymmetry breaking to the visible brane. We find that to lowest order in the five dimensional Newton's constant $k_5^2$  and gravitino mass $m_{3/2}^2$ the spinor field of the radion multiplet is responsible of inducing positive one-loop squared masses 
$m_{\varphi }^2 \sim {m_{3/2}^2} / ( M_{Planck}^2 \; R^2 )$
to the scalar fields which are localized on the visible brane with $R$ the length scale of the fifth dimension. Considering a cubic superpotential on the visible brane we also find that non-vanishing soft trilinear scalar couplings $A$ are induced given by $A=3m_{\varphi }^2/m_{3/2}$.

\end{abstract} 
\end{titlepage} 
\newpage 
\baselineskip=18pt 
\section{Introduction}
During the last years a lot of effort has been expended in the study of the physics of extra dimensions.  
Especially the assumed brane picture of our world has attracted much interest, mainly because of the new insight offering in particle physics beyond the standard model, in cosmology and the interplay between them. One of the main topics that the brane world models have been invoked for is the hierarchy problem  \cite{randall1, randall2} which is connected with the origin of the mass scale of the electroweak symmetry breaking \cite{large1, large2, arkani1}. On the other hand these models have their origin in String Theory where Supersymmetry is a basic ingredient \cite{horava, lykken, ovrut}. 
One of the main issues that may be addressed in supersymmetric brane world models is the mediation of the supersymmetry breaking and the determination of the soft-breaking terms appearing in the corresponding four-dimensional low energy theories. 
These models may be constructed by orbifolding a supersymmetric five dimensional theory with a compact extra dimension. The supersymmetry breaking is triggered on the hidden brane, which in some sense replaces the hidden sector of four spacetime dimensional models \cite{hiddensec}, and through the bulk is communicated to the visible brane \cite{peskin, bagger1, nilles, fived, lalak, moroi, radion1, gates, riotto, anomalymed1, anomalymed2}. This transmission results to  finite one-loop mass corrections for the scalar fields that live on the visible brane. The induced corrections have been already calculated in \cite{riotto, rattazzi,scrucca, gregoire, tricher} and result to tachyonic masses, although it has been claimed  that a full treatment of the radion multiplet may turn this picture yielding positive masses squared. 

In this work we study the transmission of the supersymmetry breaking in $N=2$, $D=5$ Supergravity \cite{fivesugra, gunaydin, recent} compactified on a $S_1/Z_2$ orbifold by working directly in the on-shell scheme.
The orbifolding determines two branes, the visible brane  located  at $x^5=0$ and the hidden one at $x^5=\pi R$. We construct the $N=1$ supesymmetric couplings of the brane chiral multiplets with the bulk fields and found that they are determined by a K$\ddot{a}$lher function reminiscent of the no-scale model \cite{noscale,okada}. The Lagrangian derived in this way describes the full brane-radion coupling at least to order $k_5^2$, which is adequate for our purposes. Assuming a constant superpotential on the hidden brane we calculated at one-loop the soft scalar  masses squared $m_{\varphi }^2$, induced by the mediation of the  radion multiplet. These were found to be positive. Moreover by considering a typical cubic superpotential $W$ on the visible brane we found that trilinear scalar couplings are also induced which are non-vanishing.
\section{Brane Multiplet Coupling}
In a relatively recent publication \cite{peggy1} we addressed the problem of the coupling of N=1 multiplets on the boundary branes in the context of five-dimensional N=2 Supergavity orbifolds. For chiral brane multiplets we derived the coupling to all orders in the gravitational coupling constant with the bulk gavitational fields, the graviton and the gravitino. For the derivation we worked in the on-shell scheme using N\"{o}ther procedure. 
The alternative of using off-shell formulation proves too tedious due to the fact that the theory is described by numerous auxiliary fields. Besides in the on-shell scheme the possible gaugings have been classified which is essential when one wants to promote the theory to a unified theory into which the Standard Model is embedded.
In that work we did not derive the full couplings of the radion multiplet to the brane fields. The radion multiplet propagates in the bulk but it also consists of even fields able to couple to the brane fields. The first order term in the five dimensional gravitational constant, already derived in \cite{peggy1}, was found to be  
$\frac{1}{\sqrt{6}}\; ( - J^{(\varphi ) \, \mu} + \frac{1}{2}
J^{(\chi  ) \, \mu}   )\; F_{\mu \dot{5}}^0 
$
but this by itself is not sufficient for the study of the mediation of the supersymmetry breaking from the hidden to the visible brane. 
One may  continue using  N\"{o}ther procedure in order to complete the brane with the radion multiplet couplings. However this task turns out to be cumbersome after a few steps and one may seek an alternative way to derive these terms systematically using the standard 
knowledge of N=1, D=4 Supergavity.  This is undertaken in this work. The main observation is that the 
 restriction of the radion fields $T\equiv \frac{1}{\sqrt{2}}(\;e_5^{\dot{5}} - i \sqrt{\frac{2}{3}} A_5^0\;), 
\chi ^{(T)} \equiv -\psi _5^2$ on the brane 
form a chiral multiplet having lowest order transformation laws given by
\[
\delta T = \sqrt{2} \varepsilon \chi ^{(T)}\,\, , \,\, \delta \chi ^{(T)} = i \; (\partial _{\mu }e_5^{\dot{5}} - 
i \sqrt{ \frac{2}{3}} F_{\mu5}^0 \;) \; \sigma ^{\mu}\bar{\varepsilon }\,.
\]
Note the appearence of the combination $F_{\mu5}^0 \equiv \partial _{\mu}A_5^{(0)} - \partial _5A_{\mu}^{(0)}$in this transformation law. The couplings of the radion
superfield with the other brane and gravity fields will be described by a K\"{a}hler function, 
say $ \;\mathcal{F}\; $, in the usual manner encountered in N=1, D=4 supergravity \cite{w   ittenbagger}. For convenience one can split this function as $ \;\mathcal{F}\;= \;\mathcal{N}(T, T^*)\; +
 \; \mathcal{K}(T,\varphi, T^*,\varphi^*)\; $, where the first term describes the restriction of the five dimensional supergravity on the brane, which survives  even in the absence of brane multiplets, and the second is associated with the presence of brane multiplets. From our previous analysis \cite{peggy1} we concluded that the form of the second function is 
$\mathcal{K}(T, \varphi, T^*,\varphi^*)\; \equiv \Delta _{(5)}K(\varphi,\varphi^*)$, where 
$\Delta_{(5)}\equiv e^{5}_{\dot{5}} \; \delta(x^5)$. 

For the determination of the first function $ \mathcal{N}(T, T^*)\;$ we observe that by  applying  
N\"{o}ther's approach in the most natural and plausible manner, avoiding as much as possible mathematical complexities, it turns out that the restriction of the five dimensional supergravity action on the branes does not take its familiar 4-dimensional form. In fact all terms in the gravitational part of the action involve the determinant of the five-dimensional metric $e^{(4)}e_5^{\dot{5}}$, instead of $e^{(4)}$, and besides there are no kinetic terms for the real part $e_5^{\dot{5}}$ of the scalar field $T$ and the spinor field $\psi_5^2 $ of the chiral radion multiplet. 
The same situation can be also encountered in  ordinary N=1, D=4 supergravity, where for a chiral multiplet 
$( S, f_S)$ for instance, by appropriate Weyl rescalings, $e_{\mu}^m \rightarrow e^f e_{\mu}^m$ with $e^{2f}=\sqrt{2} Re(S)$, followed by appropriate shifts in the gravitino field, one can eliminate the kinetic terms for $\;Re(S), f_S$, having as an effect the appearance of $e^{(4)} Re(S)$ instead of 
$e^{(4)}$ in the Lagrangian.
From this it becomes obvious that one needs the inverse transformations to be implemented in $N=2$, $D=5$ Supergravity Lagrangian, derived by applying N\"{o}ther's proceedure. These are given by   
\begin{eqnarray}
\hat{e}_{\mu}^m &=& e^{-f}e_{\mu}^m \nonumber \\
\hat{\psi} _{\mu}^{1,2} &=& e^{-f/2}(\psi _{\mu}^{1,2}\pm \frac{i}{2}\sigma _{\mu} \psi _5^{2,1} e^{-2f}) \nonumber \\
\hat{\psi} _{5}^{1,2} &=& e^{f/2}\psi _{5}^{1,2} \; ,  
\label{trans2}   
\end{eqnarray}
, with $e^{2f}=e_5^{\dot{5}}$,
and are able to cast the bulk Lagrangian, and especially its restriction on the branes, in the typical form reminiscent of the $N=1$, $D=4$ Supergravity in which the kinetic terms for  $e_5^{\dot{5}} $ and  $ \,\psi_5^2$ are present. In Eq. (\ref{trans2}) the hatted fields are those describing the original $N=2$, $D=5$ action. 
Then in terms of the transformed (unhatted) fields the couplings of the brane fields are  those of the N=1 , D=4 supergravity derived from the K\"{a}hler function
\begin{eqnarray}
{\cal{F}}\;= -3\;ln\frac{T+T^*}{\sqrt{2}} + \delta (x^5)\;  \frac{\sqrt{2}}{T+T^*}\;K(\varphi ,\varphi ^*).
\label{kahler}
\end{eqnarray}
At this point let us remark that in this Lagrangian one has  to substitute
$\partial _{\mu}A_5^{(0)}\rightarrow F_{\mu 5}^{(0)} $ and 
$ \partial _{\mu}\psi _5^{2}\rightarrow \partial _{\mu}\psi _5^{2}-\partial _5 \psi _{\mu}^2$ 
since these combinations actuall appear in the 5-D supersymmetric transformations of the fields.

The interaction of the brane fields with the radion multiplet stems from the following Lagrangian, where for simplicity we do not present the four-fermion terms,
\begin{eqnarray} 
\mathcal{L}_{0} &=& - e^{(4)} \mathcal{F}_{ij^*} \left[
\partial _{\mu}\varphi^i \partial ^{\mu}\varphi^{*j} +
\frac{i}{2}\left( \chi^i \sigma^{\mu} D_{\mu}\bar{\chi}^j + 
\bar{\chi}^j \bar{\sigma}^{\mu}D_{\mu}\chi^i    \right) \right] \nonumber \\
&-&i \; \frac{e^{(4)}}{4} \left[ 
\mathcal{F} _{ij^*}  \mathcal{F}_m\partial _{\mu}\varphi ^m 
-2\mathcal{F}_{mij^*}\partial _{\mu}\varphi ^m - h.c.
\right] \chi^i\sigma ^{\mu} \bar{\chi}^j \nonumber \\
&-& \frac{e^{(4)}}{\sqrt{2}} \mathcal{F}_{ij^*}\left( 
\partial _{\nu}\varphi ^{*j}\chi ^i \sigma^{\mu} \bar{ \sigma}^{\nu}\psi _{\mu} + h.c.  \right) 
+ \frac{e^{(4)}}{4} E^{\kappa \lambda \mu \nu } ( \mathcal{F}_{m} \partial _{\kappa }\varphi ^{m} - 
h.c)\; \psi _{\lambda } \sigma_{\mu} \bar{\psi }_{\nu } \; .
\label{lagrange}
\end{eqnarray}
A non-trivial superpotential $W(\varphi)$,  giving rise 
to Yukawa and potential terms, can be easily incorporated \cite{peggy1} given by 
\begin{eqnarray}
\mathcal{L}_Y + \mathcal{L}_P 
&=& - e^{(4)} \Delta_{(5)}  e^{ \;\mathcal{F}/2} ( \;
W^* \psi_{\mu} \sigma ^{\mu \nu} \psi_{\nu}  
+ \frac{i}{\sqrt{2}} D_i W \chi^i \sigma^{\mu} \bar{\psi}_{\mu} 
+ \frac{1}{2} D_i D_j W \chi^{i}\chi^{j} + h.c. \;) \nonumber \\
&&- e^{(4)} (\Delta_{(5)})^2 e^{\;\mathcal{F}} (\;  
\mathcal{F}^{ij^*}\; D_i W \; D_{j^*}W^*  - 3  \; | W |^2 \;)
\label{potential}
\end{eqnarray}
\noindent
In the Lagrangians above and in what follows $\psi_{\mu}$ stands for $\psi_{\mu}^{1}$, the even gravitino field which lives on the visible brane and the bulk as well. 
Some comments concerning the above Lagrangian are in order:
\newline
{\bf i.} The part $\mathcal{L}_0$ contains  both the terms describing the interaction of the radion multiplet with the fields localized on the branes and also terms involving only the radion multiplet fields which live in the bulk. This is due to the particular form of the K\"{a}hler metric arising from the K\"{a}hler function 
$ \mathcal{F} $ of Eq. \ref{kahler}. 
From the the bosonic and fermionic kinetic terms in the Lagrangian of Eq. (\ref{lagrange}) we get in a straightforward manner that
\begin{eqnarray}
{\cal{L}}_{kin} &= &- \frac{1}{2} \; e^{(4)}  \left(\frac{3}{2}+  \Delta_{(5)} K \right) \left[\;
\partial _{\mu} (lne_5^{\dot{5}}) \; \partial^{\mu} (lne_5^{\dot{5}}) +
 \frac{2}{3} F_{\mu \dot{5}}^{(0)} F^{\mu (0)}_{ \dot{5}} 
\right]  \nonumber \\
&&- e^{(4)}\Delta_{(5)}\left[ K_{\varphi \varphi ^*}\partial_{\mu} \varphi \partial^{\mu}\varphi ^* 
 -\frac{1}{2} \partial_{\mu} K \partial^{\mu}(lne_5^{\dot{5}}) + 
\frac{1}{\sqrt{6}}J^{(\varphi )\mu}F_{\mu \dot{5}}^{(0)} \right] \nonumber \\
&&- i \;\frac{e^{(4)}}{2} \left[ \left(\frac{3}{2} +  \Delta_{(5)} K \right)
 \psi _{\dot{5}}^2\sigma^{\mu}D_{\mu} \bar{\psi} _{\dot{5}}^2
+ \Delta_{(5)}  K_{\varphi \varphi ^*} \; \chi \sigma^{\mu}D_{\mu}\bar{\chi } -h.c.
\right] \nonumber \\
&& - i \;\frac{e^{(4)}}{2 \sqrt{2}} \; {\Delta_{(5)} }\left[ \;
K_\varphi \; (\;\chi  \sigma^{\mu}D_{\mu} \bar{\psi}_{\dot{5}}^2 + 
\bar{\psi} _{\dot{5}}^2\bar{\sigma}^{\mu}D_{\mu}\chi \;) - h.c. \right] \; \; .
\label{f5}
\end{eqnarray} 
We see that we have non-canonical kinetic terms in the bulk for the fields of the radion multiplet. From these terms only the kinetic term of the field $A_5^{(0)}$ remains if we express the action in terms of the untransformed hatted fields. Moreover we see that these terms remain on the brane multiplied by 
$\frac{2}{3}  \Delta_{(5)} K $. Cancellation  of the $\Delta_{(5)} K $ terms in the kinetic part of 
$lne_5^{\dot{5}}$ can be achieved if one adds pure gravity terms and gravitino kinetic terms localized 
on the brane. However since this is not mandatory for the N=1 supersymmetry invariance of the brane action we choose to keep the pure supergravity part in the bulk as it appears above. Notice the presence of the term 
$- \frac{1}{\sqrt{6}}J^{(\varphi )\mu}F_{\mu \dot{5}}^{(0)}$ in Eq. (\ref{f5}). An additional contribution 
$\frac{1}{2 \sqrt{6}}\;   J^{(\chi  ) \, \mu}   \; F_{\mu \dot{5}}^0 $
stems from the remaining terms of Eq. (\ref{lagrange}), \cite{peggy1}.  

{\bf ii.} In the above formulae we have written the Lagrangian for a chiral multiplet located on the brane at $x^5=0$. Similar expression holds for the hidden brane at $x^5=\pi R$. We have just to add a hidden part 
K\"{a}hler function 
$\mathcal{F}_H =  \delta (x^5-\pi R) \; \frac{\sqrt{2}}{T+T^*}\; K_H(\varphi_H ,\varphi_H ^*)$ 
 to ${\cal{F}}$ of Eq. (\ref{kahler}) depending only on the hidden brane fields and 
the corresponding superpotential $W_H$. In our work we will consider a constant superpotential on the hidden brane triggering spontaneous symmetry breaking of supersymmetry. 

{\bf iii.} The extra power of the $ \Delta_{(5)}$ prefactor multiplying the potential terms is cancelled in the first term since the inverse of the K\"{a}hler metric already includes the inverse $ (\Delta_{(5)})^{-1}$. However it is not cancelled in the second term which is proportional to $| W |^2$. Such singularities are not new and also occur in N=2 five-dimensional supersymmetries in flat space-time N=2, D=5 supersymmetries where they cure singularities arising from the propagation of the bulk fields in order to maintain supersymmetry \cite{peskin}.
We also remark that the negative term of the potential $-3 \;| W |^2$ flips its sign in the scalar potential 
if we consider the coupling with the radion multiplet. To illustrate this we consider for 
simplicity just one chiral multiplet on the visible brane with superpotential given by 
$W(\varphi ) = \frac{\lambda }{6}\varphi ^3$ and a K\"{a}hler function $K(\varphi \varphi^*) = \varphi \varphi ^*$. Then the scalar potential up to  $k_5^2$ order is found to be  
\[
\mathcal{L}_P = -e^{(4)} \Delta_{(5)}  e^{\mathcal{F}} \; \left( \;
\frac{\mid \lambda  \mid ^2}{4} \varphi^2 \varphi^{*2} + \Delta_{(5)} 
\frac{\mid \lambda \mid ^2}{12} \varphi^3 \varphi^{*3} \; \right) \; .
\]
Thus despite the fact that a negative term appears in the potential nevertheless the potential turns always positive. This feature is independent of the particular forms of $K$ and the superpotential $W$ which depend on the visible fields. This form of the potential yields the possibility of de-Sitter metastable vacua \cite{kklt}. This can be accomplished in other considerations at the cost of introducing extra D-terms in the action \cite{paraw}.

\section{Transmission of the Supersymmetry Breaking}
\subsection{Soft scalar masses }
For the purpose of studying the transition of supersymmetry breaking we consider 
a constant superpotential $c$ on the hidden brane. The corresponding Lagrangian is
\begin{eqnarray}
\mathcal{L}_h = - e^{(4)} \Delta_{(5)}^{(h)}  e^{ \;\mathcal{N}/2} ( \;
c^*  \psi_{\mu}  \sigma ^{\mu \nu} \psi_{\nu}  
+ \frac{3i}{2} c \psi _{\dot{5}}^2 \sigma^{\mu} \bar{\psi}_{\mu} 
+ \frac{3}{2} c \psi _{\dot{5}}^2 \psi _{\dot{5}}^2 + h.c. \;) 
\label{gravit}
\end{eqnarray}
where $\Delta_{(5)}^{(h)}$ is similar to $\Delta_{(5)}$ for the hidden brane. 
The absence of scalar potential terms is justified by the fact that in the absence of brane chiral multiplets the corresponding K\"{a}hler function is that of a no-scale model. 
In this case we see that mass terms for the gravitino $\psi_{\mu}$ and the spinor field $\psi _{\dot{5}}^2$ of the radion multiplet arise on the hidden brane. Moreover  the kinetic term 
$\Delta_{(5)} K
\left( \psi _{\dot{5}}^2\sigma^{\mu}D_{\mu} \bar{\psi} _{\dot{5}}^2 + 
\bar{\psi} _{\dot{5}}^2\bar{\sigma}^{\mu}D_{\mu}\psi _{\dot{5}}^2\right) 
$
appears on the visible brane. These two terms communicate through the bulk propagation 
of the spinor field of the radion multiplet yielding non-vanishing masses for the 
scalar fields of the brane chiral multiplet. We choose $K=\varphi \varphi ^*$ and for the calculation 
of the mass corrections we choose a gauge in which the bulk kinetic terms of the gravitinos are disentangled from their fifth components. That done we treat the mass terms on the hidden branes as interaction 
terms. This is sufficient for our purposes since we are interested in  mass corrections for the brane scalar fields which are of order $m_{3/2}^2 \propto {\mid c \mid}^2 $.

The diagonalization of the $\psi _{\mu}^{1} \equiv \psi _{\mu}$, $\psi _{\mu}^{2}$ and the $\psi _{\dot{5}}^{1,2}$ bulk kinetic terms is 
achieved by chosing  appropriately the  gauge fixing term. 
The five-dimensional gravitino kinetic terms expanded in components are:
\begin{eqnarray}
&& i \left( \psi _{m}^{1 } \sigma^{mnr}\partial _n  \bar{\psi }_{r}^{1 } + (1 \rightarrow 2)  \right) +
2 \left( \psi _{m}^{1 } \sigma^{mn}\partial _{\dot{5}}  \psi _{n}^{2 }  + h.c.  \right) \nonumber \\
&& -\frac{3i}{2}\left( \psi _{\dot{5}}^{1 } \sigma^{m}\partial _m  \bar{\psi }_{\dot{5}}^{1 } + (1 \rightarrow 2)  \right)  
-3 \left( \psi _{\dot{5}}^{2 }\partial _{\dot{5}}  \psi _{\dot{5}}^{1 }  + h.c  \right)  \nonumber \\
&& + \frac{3i}{2} \left( \psi _{m}^{1 } \sigma^{m}\partial _{\dot{5}}  \bar{\psi }_{\dot{5}}^{1 } +
(1 \rightarrow 2)  - h.c.   \right) 
\label{kingra}
\end{eqnarray} 
Note that  the $m , \dot{5}$ components are mixed only through $\dot{5}-$derivatives.

In our approach we employ the following gauge fixing  by adding to the bulk Lagrangian the term 
\begin{eqnarray}
i \; \frac{\xi }{2} \; \bar{\hat \Psi }_{i \tilde{m}}
 \gamma ^{\tilde{m}}  \gamma ^{\tilde{r}} 
 \gamma ^{\tilde{n}} \partial _{\tilde{r}} {\hat \Psi}_{\tilde{n}} ^i \; \;.
 \label{gfix}
\end{eqnarray}
In this ${\hat \Psi}_{\tilde{m}} ^i$, $i=1,2$, denote 4-component gravitinos in the original Lagrangian and  
$\tilde{m}=(m, \dot{5})$  are flat five-dimensional indices
{\footnote{ ${\hat \Psi}_{\tilde{m}}^{1,2}$ defined by 
$ {\hat \Psi}_{\tilde{m}}^{1}= \left( \begin{array}{c} {\hat \psi}_{\tilde{m}}^{1}  \\ 
{\bar{{\hat \psi}}_{\tilde{m}}^{2} } \end{array} \right)$ and 
$ {\hat \Psi}_{\tilde{m}}^{2}= \left( \begin{array}{c} {\hat \psi}_{\tilde{m}}^{2}  \\ 
-{\bar{{\hat \psi}}_{\tilde{m}}^{1} } 
\end{array} \right)$ are symplectic Majorana gravitinos of the N=2, D=5 Lagrangian.
}}.
Expanding this term and using the transformations of Eq. (\ref{trans2}) this can be cast in a form involving the fields $\psi_{\mu}^{1,2}, \psi_{5}^{1,2}$. 

\begin{figure}
\begin{center}
\includegraphics[width=8.5cm, height=7.3cm]{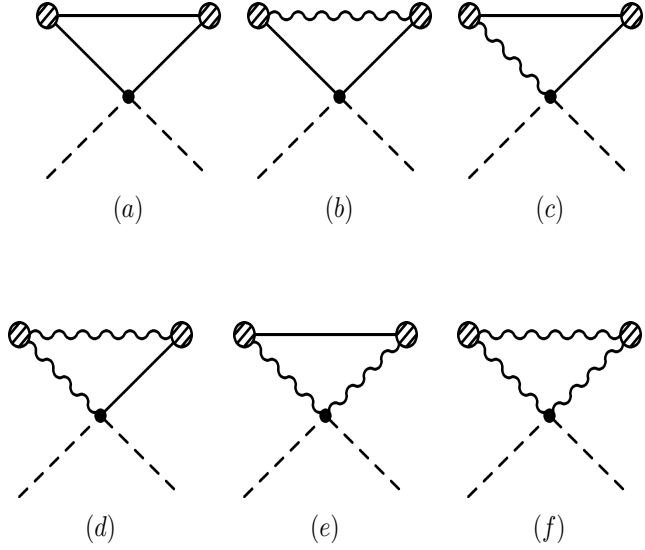}
\end{center}
\caption[]{Diagrams relevant for the calculation of the induced scalar field masses. The dashed lines denote the scalar fields lying on the visible brane. The curly lines denote the gravitinos and the solid lines stand for spinor fields of the radion multiplet. The blobs on top of each diagram are fermionic mass insertions at the hidden brane }
\label{fig1}  
\end{figure}

The gauge choice $\xi = - \frac{3}{4}$ and an additional shift implemented by 
\begin{eqnarray}
\psi _{\dot{5}}^{1,2} \longrightarrow \psi _{\dot{5}}^{1,2} \pm \frac{i}{3} \sigma ^{m} \bar{\psi}_{m}^{2,1}
\label{shift} 
\end{eqnarray}
eliminates the $m , \dot{5}$ mixings from both the kinetic part of the five-dimensional gravitinos in Eq. (\ref{kingra})  and from the terms arising from the gauge fixing in Eq. (\ref{gfix}). This results to 
\begin{eqnarray}
i \left[ \psi _{m}^{1 } \left( \sigma^{mnr} - \frac{1}{2}\sigma^{m}\sigma^{n} \sigma^{r}   \right) \partial _n 
 \bar{\psi }_{r}^{1 } + (1 \rightarrow 2) 
\right]  +
 \eta ^{mn} \left( \psi _{m}^{1 }\partial _{\dot{5}} \psi_{n}^{2 }  + h.c.  \right) \nonumber \\
-\frac{9i}{4}\left( \psi _{\dot{5}}^{1 } \sigma^{m} \partial _m  \bar{\psi }_{\dot{5}}^{1 } + (1 \rightarrow 2)  \right)  
-\frac{9}{4} \left( \psi _{\dot{5}}^{2 }\partial _{\dot{5}}  \psi _{\dot{5}}^{1 }  + h.c  \right) \; \; .
\end{eqnarray} 
The terms mixing 1 and 2 that are left over may  be further diagonalized by appropriate rotations, using four-component Majorana spinors or more conveniently by using Dirac spinors defined by 
\[
 \Psi = \left( \begin{array}{c} \psi _{\dot{5}}^{2 }  \\\bar{ \psi} _{\dot{5}}^{1 }  \end{array} \right)\,\,,
\,\, \Psi_m = \left( \begin{array}{c} \psi _m ^{1 }  \\\bar{ \psi} _{m}^{2 }  \end{array} \right) \; \; 
\]
whose upper components are constituted of even fields . 
We choose to proceed with the second choice and since we are interested in diagrams involving propagation from the hidden to the visible brane we use the pertinent Dirac and gravitino propagators in the mixed momentum-configuration space representation \cite{arkani1, puchwein, meissner}. In this representation, and in the particular gauge with the value of  $\xi$ chosen as above, the orbifolded propagators read as:
\begin{eqnarray}
&&G_{mn}(p,y,y') = \left( \frac{1}{2} \gamma _n \slash{p} \gamma _m +
 i \eta _{mn} \gamma ^{\dot{5}} \partial _y \right) \; F(p,y,y') \nonumber \\
&&G (p,y,y') = \frac{2 \;i}{9} ( \slash{p} + i  \gamma ^{\dot{5}} \partial _y ) \; F(p,y,y') 
\end{eqnarray}
where $F(p,y,y') $ is given by 
\begin{eqnarray}
&&F(p,y,y') \equiv  \frac{1}{2qsin(q \pi R)}
 \left\{
cos \left[q(\pi R - \mid y - y' \mid )\right] - i \gamma ^{\dot{5}} cos \left[q(\pi R -  y - y') \right] 
  \right \} \; \; .
\end{eqnarray}
In these $y, y'$ denote variables along the fifth dimension and $q = \sqrt{-p^2 + i\epsilon }$.
In order to proceed further we have to perform the shifts (\ref{shift}) in the relevant
terms of the brane Lagrangians. That done, the pertinent visible brane terms are brought to the form 
\[
-i \varphi \varphi ^* \left[ \bar{\Psi} P_R \slash{\partial } P_L \Psi  + 
\frac{1}{9} \bar{\Psi}_m P_R \gamma ^m  \slash{\partial } \gamma ^n P_L \Psi _n +  \frac{i}{3}(
 \Psi_m ^T P_L C \gamma ^m \slash{\partial } P_L \Psi  - h.c )\right]
\]   
while on the hidden brane we have  
\[
- \mid c \mid \left[ \Psi_m ^T P_L C( \gamma ^{mn} -\frac{1}{3}\gamma ^{m} \gamma ^{n} ) P_L \Psi_n  + 
\frac{3}{2} \Psi ^T P_L C P_L \Psi - \frac{i}{2} \bar{\Psi }_m P_R \gamma ^m P_L \Psi + h.c.  \right] 
\; \; .
\] 
In these $P_{L,R} = \frac{1}{2}(1 \pm i \gamma ^{\dot{5}})$ are the chiral projection operators and $C$ denotes the charge conjugation matrix. \\  
Calculating the  diagrams,  depicted in \ref{fig1}, which are  relevant for the scalar mass terms to order $\mid c\mid ^2$ in the supersymmetry breaking scale, we find the mass corrections to the scalar fields involved. 
The external momenta of the external scalar fields in these graphs have been taken vanishing.
The general structure of the loops involved are
\begin{eqnarray}
&&\int \frac{d^4p}{{(2 \pi)}^4} \; dy dy_1 dy_2 \; \delta(y)\; \delta(y_1-\pi R)\; \delta(y_2 - \pi R) \;
Tr \left[  V G(p, y,y_1) V_1 G(p,y_1,y_2) V_2 G(p,y_2,y) \right] \nonumber \\
&&= 
\int \frac{d^4p}{{(2 \pi)}^4}  \;
Tr \left[  V G(p, 0,\pi R) V_1 G(p,\pi R,\pi R) V_2 G(p,\pi R,0) \right] \; .\nonumber
\end{eqnarray}
In this all space-time indices have been suppresed and $V$ and $ V_{1,2}$ are vertices on the visible and hidden brane respectively. $G(p,z,z')$ denote propagations between $z$ and $z'$ points for the gravitino and fermion fields carrying the loop momentum $p$. The variables $y_{1,2}$ specify points on the hidden brane, located at $\pi R$, and $y$ those of the visible brane located at $y=0$.
The corrections  to the scalar masses squared of the various diagrams depicted in figure \ref{fig1} are of the form 
$$\Delta m_\varphi^2 = c_{(i)} \; \frac{k_5^2}{27} \; m_{3/2}^2 \; \frac{\zeta   (3)}{\pi ^5 R^3}$$ where the subscript $i$ in the constant $c_{(i)}$ labels each graph. $c_{(i)}$ are given by 
\begin{eqnarray}
c_{(a)} &=& 2 \quad , \quad
c_{(b)} = \frac{1}{4} \quad , \quad
c_{(c)} =   -1 \nonumber \\
c_{(d)} &=&  1 \quad , \quad
c_{(e)} =   \frac{1}{8} \quad , \quad
c_{(f)} =  - \frac{11}{16} \nonumber 
\end{eqnarray}
Collecting all contributions entails to the following finite mass correction
\begin{eqnarray}
m_\varphi ^2 = \frac{k_5^2}{16}\;  \frac{\zeta (3)}{\pi ^5 R^3}\; m_{3/2}^2 \; = 
\frac{\zeta (3)}{\pi^3 R^2} \; \frac{m_{3/2}^2}{M_{Planck}^2} \; .
\end{eqnarray}
In the above expressions we have reinstated the dimensions and we have made use of the fact that 
$m_{3/2} = k_5^2 \mid c \mid / (\pi R)$.
We note that the supersymmetry breaking through a constant superpotential on the 
hidden brane has resulted to non-tachyonic scalar masses for the brane fields. 

 In our approach the positivity of $m_\varphi ^2$ is intimately related with the presence of the spinor field of the radion multiplet $\psi _{\dot{5}}^2$. Since the supersymmetry breaking occurs on the hidden brane this field cannot be gauged away by a transformation from the whole Lagrangian as it would be the case for an ordinary goldstino in the  "unitary" gauge
{\footnote{Analytic treatment of the goldstino modes in other contexts has been the subject of Refs. \cite{belyaev} and \cite{benakli}.}}. 
This feature may explain the difference from other approaches where  a "unitary" gauge is adopted to set $\psi _{\dot{5}}^2 = 0$ leaving only one diagram where the gravitino is the only propagating field. Notice that the corresponding diagram (f) of figure \ref{fig1} yields negative contribution in our case as well. However in our treatment the rest of the diagrams, involving at least a $\psi _{\dot{5}}^2$ fermion, yield contributions that render scalar masses squared positive. 
\subsection{Trilinear soft scalar couplings}
\begin{figure}
\begin{center}
\vspace{0.5cm}
\includegraphics[width=12cm,, height=3cm]{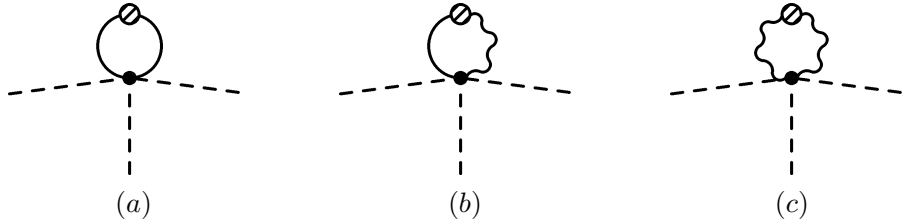}
\end{center}
\caption[]{Diagrams relevant for the trilinear soft scalar couplings. The lines are as in figure \ref{fig1}.}
\label{fig2}  
\end{figure}

For the study of the effect of the supersymmetry breaking on the trilinear scalar couplings we consider a cubic superpotential on the visible brane $W(\Phi ) = \frac{\lambda }{6}\; \Phi ^3$. The graphs one needs calculate for the trilinear couplings are shown in figure \ref{fig2}.
The necessary Lagrangian  terms for this computation stem from the Yukawa-type terms in the action which read
\begin{eqnarray}
- W(\varphi )\left[ \bar{\Psi}_m ^T P_R C( \gamma ^{mn} -\frac{1}{3}\gamma ^{m} \gamma ^{n} ) P_R C \bar{\Psi}_n^T  + 
\frac{3}{2} \Psi ^T P_L C P_L \Psi - \frac{i}{2}\bar{ \Psi}_m  P_R \gamma ^m P_L \Psi + h.c.  \right] .
\end{eqnarray}
To order $m_{3/2}$ the separate diagrams depicted in figure \ref{fig2} yield trilinear scalar field corrections given by, 
$$c_{(i)}\; m_{3/2}k_5^2 \; \frac{ \zeta (3) \; }{ \pi ^5 R^3} \; W(\varphi) + h.c.$$ 
where the coefficient $c_{(i)}$  for each graph involved is given below, 
\begin{eqnarray}
c_{(a)} =  \frac{1}{12} \quad , \quad
c_{(b)} = \frac{1}{48} \quad , \quad
c_{(c)} =  \frac{1}{12} \nonumber 
\end{eqnarray}
Adding the separate contributions one gets a correction to the cubic potential due to the supersymmetry breaking that occured on the hidden brane given by 
\begin{eqnarray}
  \frac{ 3 }{16} \;m_{3/2} \;  k_5^2\; \frac{  \zeta (3)}{ \pi ^5 \;R^3} \; W(\varphi) + h.c. \; \; .
\end{eqnarray}
Therefore the induced trilinear soft scalar coupling is $A=3m_{\varphi }^2/m_{3/2}$. 
\section{Discussion}
In the context of $D=5$, $N=2$, Supergavity compactified on 
$S^1/Z_2$ we  considered  the $N=1$ supersymmetric couplings of matter localized 
on one of the branes of the orbifold~. The inclusion of the radion multiplet couplings is accomplished  to second order in the five-dimensional gravitational constant $k_5$ working directly in the on-shell formalism. We studied the transmission of the supersymmetry breaking occuring on the hidden brane to the visible sector of the theory. In particular up to second order in $m_{3/2}$ we calculated the one-loop masses  induced to the scalar fields on the visible brane. Proper treatment of the radion multiplet shows that this transmission results to positive squared universal masses $m_{\varphi }^2 > 0$. Furthermore we found that a universal trilinear soft scalar coupling is induced due to the transmision of supersymmetry breaking given by 
$A=3m_{\varphi }^2/m_{3/2}$  which is non-vanishing and positive. These results can be easily extended in gauged supergravities.
\section*{Acknowledgements}
This work is co-funded by the European Social Fund  
and Natural Resources - EPEAEK B - PYTHAGORAS. The authors acknowledge also partial support from the Athens University special research account.  

\end{document}